\documentclass[aps,preprint,12pt,onecolumn,superscriptaddress]{revtex4-1}

\usepackage{amsmath}
\usepackage{amssymb}
\usepackage{amstext}
\usepackage{bm}
\usepackage{graphicx}
\usepackage{array}
\usepackage{multirow}
\usepackage{lipsum}
\usepackage{color}
\usepackage{ulem}
\usepackage{mathtools}
\usepackage{float}

\newcommand{\ket}[1]{\ensuremath{\left|#1\right\rangle}}

\DeclareMathOperator*{\argmax}{argmax}
\DeclareMathOperator*{\argmin}{argmin}

\begin{document}

\title{Single-Step Shaping of the Orbital Angular Momentum Spectrum of Light}
\author{Jonathan Pinnell}
\affiliation{School of Physics, University of the Witwatersrand, Johannesburg 2000, South Africa}
\author{Valeria Rodr\'{i}guez-Fajardo}
\affiliation{School of Physics, University of the Witwatersrand, Johannesburg 2000, South Africa}
\author{Andrew Forbes}
\affiliation{School of Physics, University of the Witwatersrand, Johannesburg 2000, South Africa}
\affiliation{Andrew.Forbes@wits.ac.za \vspace*{1cm}}

\begin{abstract}
\noindent Control of orbital angular momentum (OAM) in optical fields has seen tremendous growth of late, with a myriad of tools existing for their creation and detection. What has been lacking is the ability to arbitrarily modify the OAM spectrum of a superposition in amplitude and phase, especially if \textit{a priori} knowledge of the initial OAM spectrum is absent. Motivated by a quasi-mapping that exists between the position and OAM of Laguerre-Gaussian modes, we propose an approach for a single-step modulation of a field's OAM spectrum. We outline the concept and implement it through the use of binary ring apertures encoded on spatial light modulators. We show that complete control of the OAM spectrum is achievable in a single step, fostering applications in classical and quantum information processing that utilise the OAM basis.
\end{abstract}

\maketitle

\section{Introduction}

It has been known for a long time that light can carry angular momentum in the form of spin (polarization) and orbital (OAM) components, but it wasn't until approximately 25 years ago that OAM carrying fields could be easily created in the laboratory \cite{Allen92}. Since then, many approaches have been used to create structured light fields that contain scalar or vector superpositions of OAM \cite{Forbes2016,SPIEbook}, driven by the many applications such fields have cultivated \cite{roadmap2,OAM1}. In a similar vein, the detection of OAM has likewise received much attention with deterministic approaches based on conformal mapping solutions in so-called mode sorters \cite{Berkhout2011A,Ruffato2018}. What has remained somewhat lacking is the ability to arbitrarily modify some initial (perhaps unknown) OAM spectrum in amplitude and phase. So far, only limited tools exist based primarily on the principle of exploiting some property such as parity (odd/even modes) and symmetry (symmetric/antisymmetric modes) and routing the resulting OAM into paths \cite{Berkhout2010,Lavery2011A,zhang2016engineering,liu2016demonstration}. But such approaches have not yet been able to shape the OAM spectrum because of the complexity and the limited versatility of the techniques. Further, once separated, each path would have to be appropriately modified and then recombined again later through an inverse system, thus requiring a multi-element and multi-step process. Often such approaches employ interferometers, further reducing the up-take due to stability issues. An approach to modify the OAM spectrum directly has been to control the range of the field's azimuthal angle (the conjugate Fourier variable of OAM) \cite{Jack2008A}, however this mostly only affects the dispersion of the OAM spectrum. Similarly, it was shown that phase dislocations, discontinuities and delays affect the OAM spectrum in distinct ways which has applications in imaging \cite{Torner2005Digital}, but a method of tailoring the OAM spectrum in this way has not been demonstrated. 

Altogether, arbitrary shaping of the OAM spectrum of light remains an open challenge. Yet, many applications require this, for example, the quantum Fourier transform as well as crucial steps in quantum computational algorithms such as that of Shor \cite{shor1994algorithms} and Grover \cite{perez2018first} and general quantum information processing simulated with classical light \cite{konrad2019quantum,forbes2019classically,toninelli2019concepts}. In general, the task is to take an arbitrary initial OAM spectrum (specified by the coefficients $c_\ell$) and modify it according to $\sum c_\ell \ket{\ell} \rightarrow \sum \tilde{c}_\ell \ket{\ell}$. For instance, a key step in the Shor algorithm is to execute the transformation $c_\ell \rightarrow c_\ell \exp(2 \pi i a^\ell/N)$ for each $\ket{\ell}$ in the initial OAM spectrum where $N$ is the number one wishes to find the factors of and $a<N$. Presently, there is no protocol for modifying an arbitrary initial OAM spectrum in such a manner and in one step. One should note the distinction between generating an engineered/tailored OAM spectrum (as in \cite{Li2018engineeredOAM}) versus shaping an OAM spectrum that was previously generated. 

Here, we outline and demonstrate an approach to shaping the OAM spectrum in a single step. We circumvent the difficulties of manipulating the spectrum directly in OAM space by recognizing a quasi-mapping between the OAM and position spaces for Laguerre-Gaussian (LG) and Perfect Vortex (PV) beams, where each $\ell$ maps to a radial position $r_\ell$. Such a mapping enables the manipulation of the OAM content through manipulation of the light field at specific spatial positions, facilitating easy OAM spectrum modulating in amplitude and phase using conventional devices such as spatial light modulators (SLMs). We then implement our scheme experimentally with PV beams and show arbitrary conversion of an initial OAM spectrum to some desired spectrum. In what follows, we give theoretical and experimental credibility to such a scheme. 

\section{Concept}
We begin by phrasing the problem we wish to solve. Suppose we are given an arbitrary OAM-containing light field $U(r,\phi)$, where $(r,\phi)$ are transverse cylindrical coordinates. We can expand this field as,
\begin{equation} \label{eq:spectrum}
U(r,\phi) = \sum_\ell c_\ell\, A_\ell (r) \, \exp(i\ell\phi) \equiv \sum_\ell c_\ell \ket{\ell} \,,
\end{equation}
where $\exp(i\ell\phi)$ is an OAM eigenstate with $\ell \hbar$ of OAM per photon in the field and $A_\ell(r)$ is the radial amplitude function for the chosen set of spatial vortex modes (Laguerre-Gauss or Bessel-Gauss beams for example) which will depend on $\ell$ in general. The complex coefficients $c_\ell$ specify the OAM spectrum of this field with normalisation $\sum_{\ell} |c_\ell|^2 = 1$. Stated precisely, the problem of OAM spectral shaping is firstly to find a method of modifying/shaping the spectrum coefficients $c_\ell \rightarrow \tilde{c}_\ell = f(\ell)\, c_\ell$, where $f(\ell)$ may be any complex transformation function and secondly, to find an optical device that could implement this transformation.

\begin{figure*}[ht] 
	\centering
	\includegraphics[width=\textwidth]{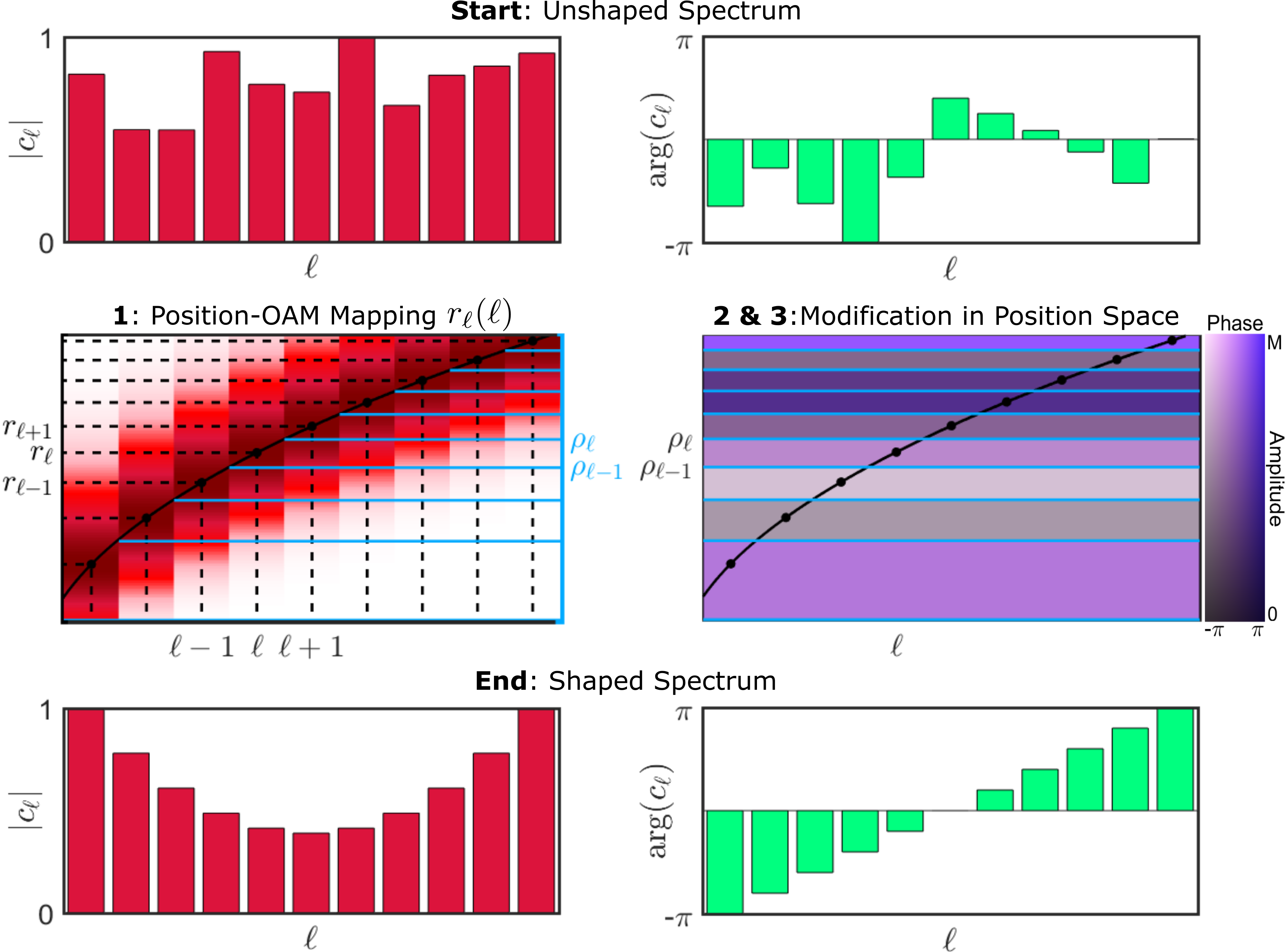}
	\caption{Concept behind our proposed OAM shaping method. Step 1 involves the determination of the position-OAM mapping ($r_\ell \leftrightarrow \ket{\ell}$) denoted by the solid line and is performed theoretically (or numerically) beforehand. Step 2 corresponds to the construction of the ring apertures centred at $r_\ell$ and having a certain width $\rho_\ell - \rho_{\ell-1}$, denoted by the horizontal bands (this is also performed theoretically beforehand). Then, by specifying the amplitudes and phases of the ring apertures (step 3) and applying them collectively to the original field, the desired shaping is elicited.}
	\label{fig:concept}
\end{figure*}

Our proposed solution to the problem as stated above is a general procedure which is outlined visually in Figure \ref{fig:concept}. Firstly, each OAM mode $\ket{\ell}$ is associated to a unique spatial (radial) position $r_\ell$ via a quasi-mapping. In general, this mapping will be unique for the chosen set of vortex modes: $A_\ell(r)$. This allows $f(\ell)$ to be replaced by $f(r_\ell)$. Since only a spatially varying function is needed, many optical devices could implement this transformation, such as spatial light modulators. Secondly, annular ring apertures centred at $r_\ell$ and of a certain width $\rho_\ell - \rho_{\ell-1}$ (to be defined later) are assigned to each OAM mode one wishes to shape. Lastly, one specifies a transmission function $t_\ell$ for each ring aperture which will be directly related to the ``amount" by which the corresponding OAM spectrum coefficient is modified ($t_\ell \sim f(\ell) $). The collection of ring apertures can then be applied to the field as a whole so as to shape the OAM spectrum in a single shot. Note that this scheme only requires the field's inherent scale parameter as input. In the following, we will describe the specifics of each step of the process in detail. 

\subsection*{Position-OAM Mapping}
As stated earlier, directly manipulating the OAM of light in an arbitrary way is difficult, particularly if the OAM spectrum is unknown. Instead, we exploit a mapping from OAM to position space $\ket{\ell} \leftrightarrow r_\ell$, as depicted by the solid line in step 1 of Fig.~\ref{fig:concept}. By way of example, consider the Laguerre-Gaussian modes $\text{LG}_p^\ell(r)$ which, at the waist plane $z=0$, are given by
\begin{align}
\text{LG}_p^\ell(r) = \sqrt{\frac{2 p!}{\pi(p + |\ell|)!}} \, \frac{1}{w_0} \, \left(\frac{r\sqrt{2}}{w_0} \right)^{|\ell|} \, L_p^{|\ell|}\left( \frac{2 r^2}{w_0^2} \right) \exp \left( -\frac{r^2}{w_0^2} \right) \,,
\end{align}
where $p$ is a radial index, $w_0$ is the Gaussian waist radius and $L_p^{\ell}(\cdot)$ are the associated Laguerre polynomials. Since $\text{LG}_p^\ell(r) = \text{LG}_p^{-\ell}$, for brevity we consider $\ell\geq 0$ and so the analysis will apply equally to $\ket{\ell}$ and $\ket{-\ell}$. Setting the radial index to zero, the radius where $\text{LG}_0^{\ell}(r)$ is a maximum is,
\begin{equation} \label{eq:LGmap}
     r_\ell = w_0 \sqrt{\frac{\ell}{2}} \,.
\end{equation}
This defines a radius at which the ring of light corresponding to the vortex beam $\ket{\ell}$ is found. Although this is a well-known result, we propose that the above represents a quasi-mapping from OAM to position space; ``quasi" in the sense that not all of $\ket{\ell}$ is contained in a ring of infinitesimal thickness at $r_\ell$ as the vortex ring has a finite width. However, by definition, most of $\ket{\ell}$ is spread in the vicinity of $r_\ell$. Now, the desired OAM transformation function $f(\ell)$ becomes a spatial transformation function $f(2 r_\ell^2 / w^2_0)$. For example, $f(\ell) \propto \ell^2$ becomes $f(r_\ell) \propto 4 r_\ell^4/w_0^4$. Note that no knowledge of the initial spectrum is required for applying this spatially dependent function.

Analogous quasi-mappings can be found for other vortex mode sets (numerically if not analytically). Since the vortex ring of light is typically the global maximum of the field amplitude, for many vortex modes the position-OAM mapping can be found from
\begin{equation} \label{eq:PosOAM}
    \ket{\ell} \leftrightarrow r_\ell = \argmax_r |A_\ell(r)| \,.
\end{equation}

\subsection*{Construction of Annular Ring Apertures}
The next step in the procedure is to construct the annular ring apertures centred at $r_\ell$, the boundaries of which are depicted by the horizontal lines in steps 2 and 3 of Fig.~\ref{fig:concept}. It seems reasonable to assemble $\ket{\ell}$'s annular ring about $r_\ell$ in such a way that it contains as much of the associated mode as possible without overlapping with adjacent rings. One has a certain degree of freedom here in choosing how to achieve this. One possibility is to choose the boundaries of the annular ring to be where the given mode $\ket{\ell}$ first intersects with the adjacent modes: $\ket{\ell \pm 1}$. In the case of LG modes, for example, the point of intersection of $\ket{\ell}$ with $\ket{\ell+1}$ can be calculated analytically (with all modes normalised to unit amplitude) and is given by,
\begin{equation} \label{eq:ringBound}
    \rho_{\ell} = \frac{w_0}{\sqrt{2 e}} \, \ell^{-\ell/2} \, \left( 1+ \ell \right) ^{\,(1+\ell)/2} \,.
\end{equation}
Using the above, one can then construct a set of ring apertures for all the OAM modes one wishes to shape. An analogous result can be found for other vortex mode sets by solving,
\begin{equation} \label{eq:GenRingBound}
    \rho_\ell = \argmin_r \, |A_{\ell}(r)| - |A_{\ell+1}(r)| \,.
\end{equation}

\subsection*{Specifying the Ring Aperture Transmission Function}
Once Eqs.~\ref{eq:PosOAM} and \ref{eq:GenRingBound} are determined for the specific vortex modes chosen, all that is left is to specify the transmission function $t_\ell$ for each ring aperture which is directly related to the desired transformation function $f(\ell)$. Physically speaking, we modify the amount and phase of light within the annular rings which, owing to the position-OAM mapping, modifies the amplitude and phase of the corresponding OAM modes. This is shown in steps 2 and 3 in Fig.~\ref{fig:concept} where the gray level within each ring aperture corresponds to $|t_\ell|$ and the colour corresponds to $\arg(t_\ell)$. For example, if one wishes to lower the contribution of $\ket{\ell}$ to the OAM spectrum relative to the others, we encode $|t_\ell| < 1$. If one wishes to apply an inter-modal phase shift to $\ket{\ell}$, we encode $\arg(t_\ell) \in [-\pi,\pi]$. 

By encoding the ring apertures on phase-only SLMs, the implementation of this step is straightforward: the amount of light within an annular ring can be modified by encoding a diffraction grating of variable height, thus displacing a controllable amount of light away from this region. Similarly, the phase of light within an annular ring can be modified by encoding a constant phase relative to some reference mode. In this implementation, light can only be removed from annular regions and so this imposes the restriction $|t_\ell| \leq 1$.

Although it's clear that the transmission function of the ring apertures and the transformation function are related, the main factor which determines their exact relationship is the OAM mode overlap. In general, OAM modes in the field will have some spatial overlap and as a consequence, this will inhibit the ability to modify each spectrum coefficient independently. The higher the spatial overlap, the more coupled the OAM modes become and the more difficult it is to determine the relationship between $t_\ell$ and $f(\ell)$ in our scheme. Conversely, if the OAM modes in the field are sufficiently spatially distinct, then $t_\ell = f(\ell)$, which is the ideal scenario.

\section{Simulated OAM Spectral Shaping}
\begin{figure*}[h!] 
	\centering
	\includegraphics[width=0.67\textwidth]{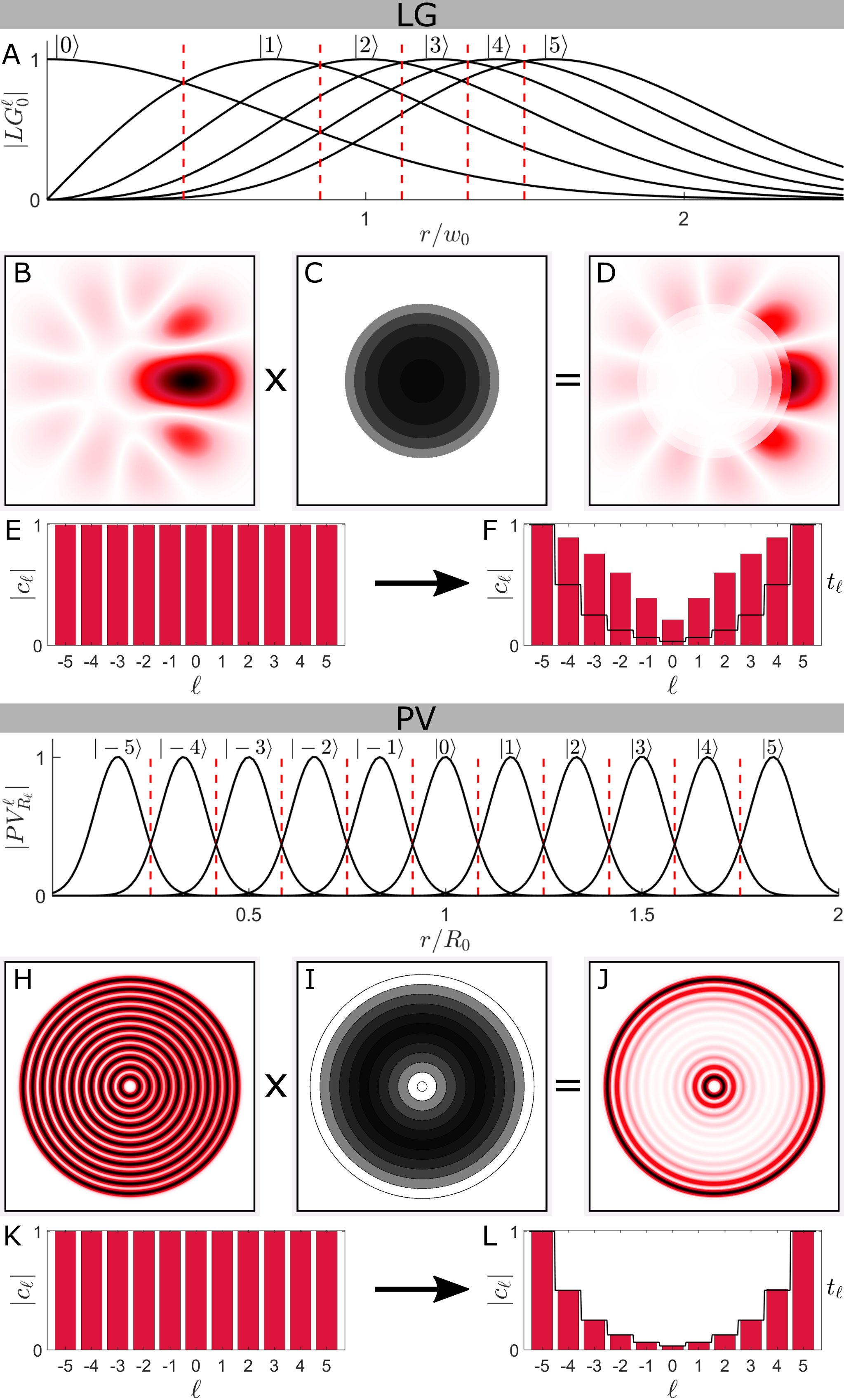}
	\caption{Plot showing the OAM mode shaping process with LGs (top) and PVs (bottom) for an initially uniform OAM spectrum. The radial field profile of individual modes is shown in A,G with vertical dashed lines denoting the boundaries of the ring apertures. The initial field, the corresponding modification hologram and the shaped field is shown in B-D and H-J. The OAM spectrum before and after shaping is shown in E,F and K,L where the solid lines in F,L denote the encoded transmittance of the ring apertures.}
	\label{fig:modeOverlap}
\end{figure*}

We can test the proposed OAM shaping scheme by simulating the generation, modification and detection of superpositions of OAM modes. An example of such a simulation for LG modes is shown in Fig.~\ref{fig:modeOverlap} A-F. Here, the following uniform ($c_\ell=c=1/\sqrt{11}$) OAM spectrum was generated
\begin{equation}
U(r,\phi) = c \sum_{\ell=-5}^5 LG_0^\ell(r) \, \exp(i\ell\phi) = c \sum_{\ell=-5}^5 \ket{\ell} \,.
\end{equation}
The transverse amplitude of this field is displayed in Fig.~\ref{fig:modeOverlap}B. As a working example, the transmittance of each ring aperture was set according to,
\begin{equation}
    t_\ell \propto 2^{\ell} \,. 
\end{equation}

Temporarily and for simplicity, we only consider the shaping of the amplitude of $c_\ell$. The collection of ring apertures (constructed using Eqs.~\ref{eq:LGmap} and \ref{eq:ringBound}) is shown in Fig.~\ref{fig:modeOverlap}C and the field amplitude immediately after modification is shown in Fig.~\ref{fig:modeOverlap}D. The measurement of the OAM spectrum was simulated by modulating the input field with a set of conjugate helical phases and observing the on-axis intensity in the far-field (a common experimental modal decomposition procedure \cite{Schulze2013}). The spectrum before and after shaping is shown in Figs.~\ref{fig:modeOverlap}E and F where we see that the shaping resulted in the transformation function $ f(\ell) \propto \sqrt{\ell}$. This demonstrates the feasibility of using ring apertures for OAM spectral shaping with LG modes. However, a complication is that it is difficult to predict exactly how the spectrum will be modified given only $t_\ell$.

As mentioned earlier, the spatial overlap between adjacent LG-OAM modes (as can be seen in Fig.~\ref{fig:modeOverlap}A) is the culprit of this unpredictability. In fact, in the limit that $\ell\rightarrow \infty$, adjacent LG-OAM modes overlap one another completely. Hence, modifying the amplitude and phase in one annular region affects the amplitude and phase of other OAM modes in a complex way. This means that unless some explicit relationship can be found, one may have to resort to trial and error to find appropriate transmission functions that will enable the intended shaping. For our example, it turns out that if we wanted $f(\ell) \propto 2^\ell$ for modes up to $\ell = 5$, we would have to encode $t_\ell \propto 2^\ell$ for ring apertures up to $r_{20}$.

We expect that the OAM shaping for other vortex modes (such as hypergeometric-Gauss and Bessel-Gauss) will exhibit similar behaviour to LGs since the OAM modes are distributed in a similar way. However, we can ask if there is a set of vortex modes that are sufficiently spatially separated to minimise modal overlap and thus allow a more predictable implementation.

In this regard, we turn to the so-called Perfect Vortex (PV) modes which are topical in the optical community owing to an annular field structure whose width is OAM independent \cite{Ostrovsky13}. We'll see that if we utilise PVs as the OAM carriers, we can engineer an OAM-containing field in such a way that the OAM modes are sufficiently non-overlapping to enable $t_\ell = f(\ell)$. Further, there is the freedom to place OAM modes in any desired order and so one is not restricted to the inherent hierarchy in ``regular'' vortex modes of having larger OAM values further from the optical axis. As an added benefit, one can now differentiate between $\ket{\ell}$ and $\ket{-\ell}$ and shape these two modes independently.


PVs are not particularly exotic; in fact, they are known to be the Fourier transform of the well-studied Bessel-Gauss (BG) beams,
\begin{equation} \label{eq:BG}
    BG_{k_r}^\ell(r) \propto J_\ell(k_r r) \, \exp\left( - \frac{r^2}{w_0^2} \right) \,,
\end{equation}
where $k_r$ is the radial wavenumber of light, $w_0$ is the Gaussian waist radius and $J_\ell(\cdot)$ is the Bessel function of order $\ell$. As such, the electric field of PV beams can be derived by Fourier transforming Eq.~\ref{eq:BG} and is described by \cite{vaity2015perfect},
\begin{align} \label{eq:PV}
    PV_{R}^\ell(r) \propto \frac{w_0}{T} \exp\left( - \frac{r^2 + R^2}{T^2} \right) I_\ell \left( \frac{2 r R}{T^2} \right)  \,,
\end{align}
where $I_\ell(\cdot)$ is the modified Bessel function, $T,R$ are respectively the annular ring thickness and ring radius
\begin{align}
    T &= \frac{2 f}{k w_0} \label{eq:rt} \,,\\
    R &= \frac{k_r f}{k} \label{eq:rr} \,,
\end{align}
and $f,k$ are the focal length of the Fourier lens and the light's wavenumber. The ``perfectness" of PVs relies on the fact that the ring thickness $T$ and ring radius $R$ are $\ell$-independent.


The basic idea for engineering non-overlapping OAM modes with PVs is to assign a single OAM mode to a single PV and to separate each PV from one another. Since PVs are generated from BGs, $k,w_0$ and $f$ will typically be global constants for all modes. This is because it is difficult to engineer a superposition of BGs where each mode has a different wavelength, width or Fourier lens. Hence, by Eqs.~\ref{eq:rt} and \ref{eq:rr}, the PV ring thickness is fixed and the ring radius is controlled by $k_r$. To force the PVs to be non-overlapping, the minimal-overlap condition can be phrased as follows: the difference in ring radii between adjacent rings should not be less than twice the ring thickness, or equivalently $\Delta R \geq 2 T$. Substituting Eqs.~\ref{eq:rt} and \ref{eq:rr}, this condition is reduced to,
\begin{equation} \label{eq:minOverlapCond}
    \Delta k_r \geq \frac{4}{w_0} \, \, \leftrightarrow \, \, \Delta R \geq \frac{4 f}{k w_0} \,.
\end{equation}
Any choice of ring spacing satisfying this inequality will enable one to modify the OAM modes independently. In what follows we choose $\Delta k_r = 6/w_0$ to make the separation of rings more apparent. Deriving a closed form expression for the intersection of $\ket{\ell}$ with $\ket{\ell+1}$ for PVs (analogous to Eq.~\ref{eq:ringBound}) is elusive. However, when $R\gg T$ (or equivalently when $k_r w_0 \gg 1$), the intersection point is approximately the midpoint of the two ring radii.


We now showcase (through simulation) the OAM spectrum shaping with PVs, which is shown in Fig.~\ref{fig:modeOverlap} G-L. Similar to the flat LG-OAM spectrum shown earlier, we consider the following superposition,
\begin{equation} \label{eq:PVsuper}
    U(r,\phi) = c \sum_{\ell = -5}^{5} PV_{R_\ell}^\ell(r) \, \exp(i\ell\phi) \,,
\end{equation}
where $R_\ell = R_{0} + \ell\Delta R$ is the OAM-position mapping with $R_0$ being the ring radius of some reference mode ($\ket{0}$ in our case) and $\Delta R$ satisfying Eq.~\ref{eq:minOverlapCond}. This field amplitude is shown in Figure \ref{fig:modeOverlap}H where each PV in the superposition is normalised to unit amplitude. The vertical dashed lines in Figure \ref{fig:modeOverlap}G denote the boundaries between PV-OAM modes. As we see, the approximation that the intersection point is the midpoint of the two ring radii is good enough in this case. The ring apertures are encoded with the same transmission functions as to those used earlier for LG modes (only that now $\ket{\ell}$ and $\ket{-\ell}$ are distinguishable). 

We observe from the shaped coefficients that the transformation function $f(\ell)$ is precisely the encoded transmittance of the ring apertures $t_\ell$ (as desired). Furthermore, the experimenter now has the freedom to choose how to structure the OAM-position mapping as it is no longer imposed by nature. In our example, the OAM-position space mapping is linear and of the form 
\begin{equation}
    \ket{\ell} \leftrightarrow R_\ell = R_0 + \ell \Delta R  \,.
\end{equation}
Note that both $R_0$ and $\Delta R$ can be manipulated as desired. It is also conceivable to manufacture a non-linear mapping.

\section{Experimental OAM Spectral Shaping}
\begin{figure*}[b] 
	\centering
	\includegraphics[width=\textwidth]{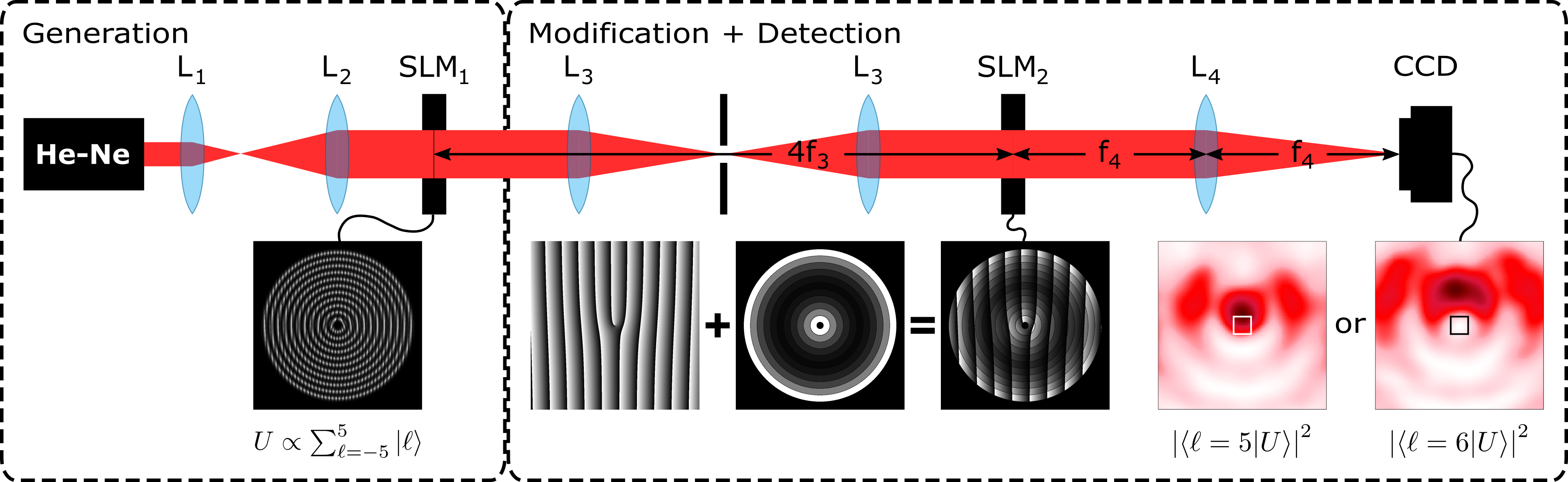}
	\caption{Schematic of the  experimental set-up; $L_i$ are lenses of focal length $f_i$. The left side shows the generation part of the setup with the corresponding hologram that creates the PV superposition mode. The right side depicts the modification and detection steps with the corresponding hologram (which is a product of modification and detection holograms) and 2 example images at the detection plane (small squares denote the on-axis region).}
	\label{fig:setup}
\end{figure*}

We now present the experimental setup used to verify the OAM shaping scheme with PVs. Similar to the simulations performed earlier, the experiment comprises generation, modification and detection phases. To this end, we built the experimental setup shown schematically in Figure \ref{fig:setup}. The generation step of the setup creates a superposition of PV-OAM modes from an expanded and collimated He-Ne laser beam using a Spatial Light Modulator (Holoeye Pluto) employing complex amplitude modulation \cite{arrizon2007}. Using a 4f lens system, this field is then relayed to a second SLM which performs the modification and detection steps simultaneously (otherwise a third SLM would be required). This can be done since these steps occur at the same plane and the transmission functions of the holograms used for modification and detection are linear.

In order to measure the OAM spectrum coefficients $c_\ell$, we utilise a standard modal decomposition setup which has been shown to be effective in detecting the OAM of PVs \cite{Pinnell:19}. Specifically, to measure $|c_\ell|$ (we do not measure inter-modal phases here for brevity) the detection system needs to perform the following overlap integral,
\begin{align}
|c_\ell| &= \left| \int_0^\infty r \,dr \int_0^{2\pi} d\phi \, U(r,\phi) \, \exp(-in\phi) \right| \,, \\
&\propto \left|\int_0^\infty r\, dr A_\ell(r) \right| \delta_{\ell,n} \equiv |\gamma_\ell| \, \delta_{\ell,n} \,, \label{eq:overlap}
\end{align}
where $\delta_{\ell,n}$ is the Kronecker delta symbol, $\exp(-i n\phi)$ is an OAM eigenstate and $\gamma_\ell$ is the value of the radial integral. Ideally, there would be no $\ell$-dependent pre-factor ($|\gamma_\ell|$) for the Kronecker delta in Eq \ref{eq:overlap}, but since the radial amplitude is $\ell$-dependent (in general) the radial integral will be too. The overlap integral can be performed optically by modulating the field with a spiral phase optic whose transmission function is $\exp(-in\phi)$ (such as a forked hologram displayed on a SLM), passing the modified field through a Fourier lens and measuring the on-axis intensity at the Fourier plane using a camera. The Kronecker delta in Eq.~\ref{eq:overlap} means that we will measure an on-axis intensity if and only if the charge of the forked hologram matches an OAM mode in $U(r,\phi)$, with the on-axis intensity being proportional to $|c_\ell|^2$. In this way, the OAM spectrum coefficients are obtained by scanning through a set of forked holograms and measuring the on-axis intensity. One can also choose to multiplex many forked holograms having different diffraction gratings and perform the full set of OAM measurements in a single shot. 

\subsection*{Correction Factors}
Equation \ref{eq:overlap} suggests that the raw spectrum we measure requires post processing in the form of correction factors (this is true more generally when decomposing into a non-orthonormal basis such as the OAM eigenstates). Experimentally, $|\gamma_\ell|^2$ controls the magnitude of the on-axis intensity that we measure. For the case of PVs, it turns out that $|\gamma_\ell|^2$ is proportional to the energy of each PV annulus in the transverse plane. Since each PV mode in the superposition in Eq.~\ref{eq:PVsuper} has the same thickness but a different radius and thus a different area, the transverse energy of each OAM carrier is different. This means that we would measure different on-axis intensities for different OAM modes even though they may have the same weight in the spectrum. Hence, before shaping can be performed, we should ensure to normalise the energy in each OAM mode so as to measure the true spectrum. One way to do this is to apply the factors $1/|\gamma_\ell|$ to the raw coefficients $|c_\ell|$. These factors can be determined in one of two ways: numerically or empirically.

\begin{figure*}[b]
\centering
\includegraphics[width = \textwidth]{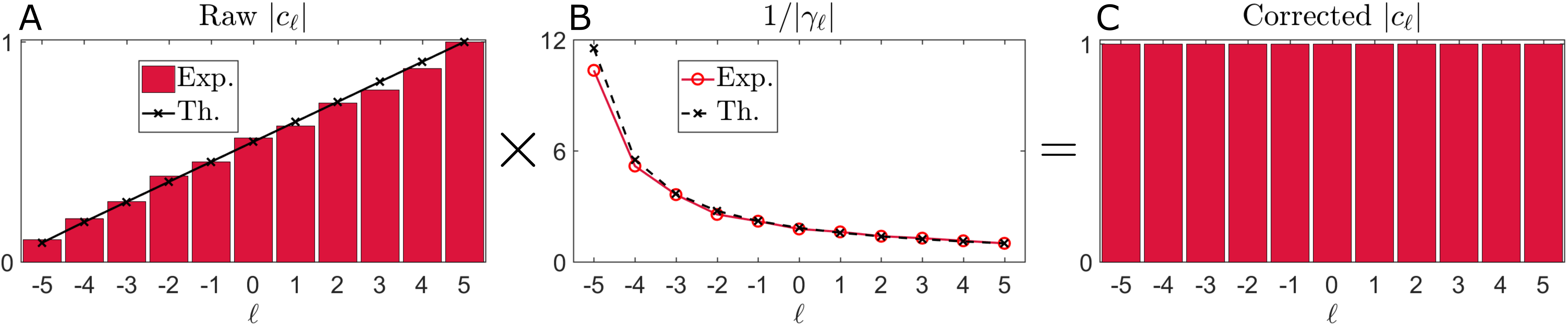}
\caption{Procedure for OAM spectrum correction. Shown from left to right are the experimental and theoretical raw spectra coefficients, correction factors and corrected coefficients (respectively) for the decomposition of the field in Eq.~\ref{eq:PVsuper}. Raw spectra coefficients are normalised to the maximum camera gray level for the particular settings used.}
\label{fig:corrections}
\end{figure*}

To numerically correct the measured on-axis signal we compute $|\gamma_\ell|$ explicitly from the radial integral in Eq.~\ref{eq:overlap}. These values correspond to the dashed line Figure \ref{fig:corrections}B. This approach to correcting the spectrum will work assuming the generation and detection system is already well aligned and calibrated.

Empirically correcting the raw OAM spectrum involves sending a uniform superposition through the detection system. The correction factors are then precisely the inverse of each measured spectrum coefficient. This is also an effective way to correct for small systematic errors in the generation and/or detection system, such as a non-uniform beam from the laser, errors in determining the on-axis camera pixel coordinates or misalignment of the detection hologram. Both types of corrections are shown in Figure \ref{fig:corrections}B. We see that the numerical and empirical corrections are in good agreement, indicating that our generation and detection systems are well aligned and calibrated. In our experiment, the source of systematic error was the fact that the laser beam at SLM$_1$ had some residual Gaussian intensity variation, resulting in modes near the optical axis having slightly more power than they should.

\subsection*{Results}
\begin{figure*}[ht]
\centering
\includegraphics[width = \textwidth]{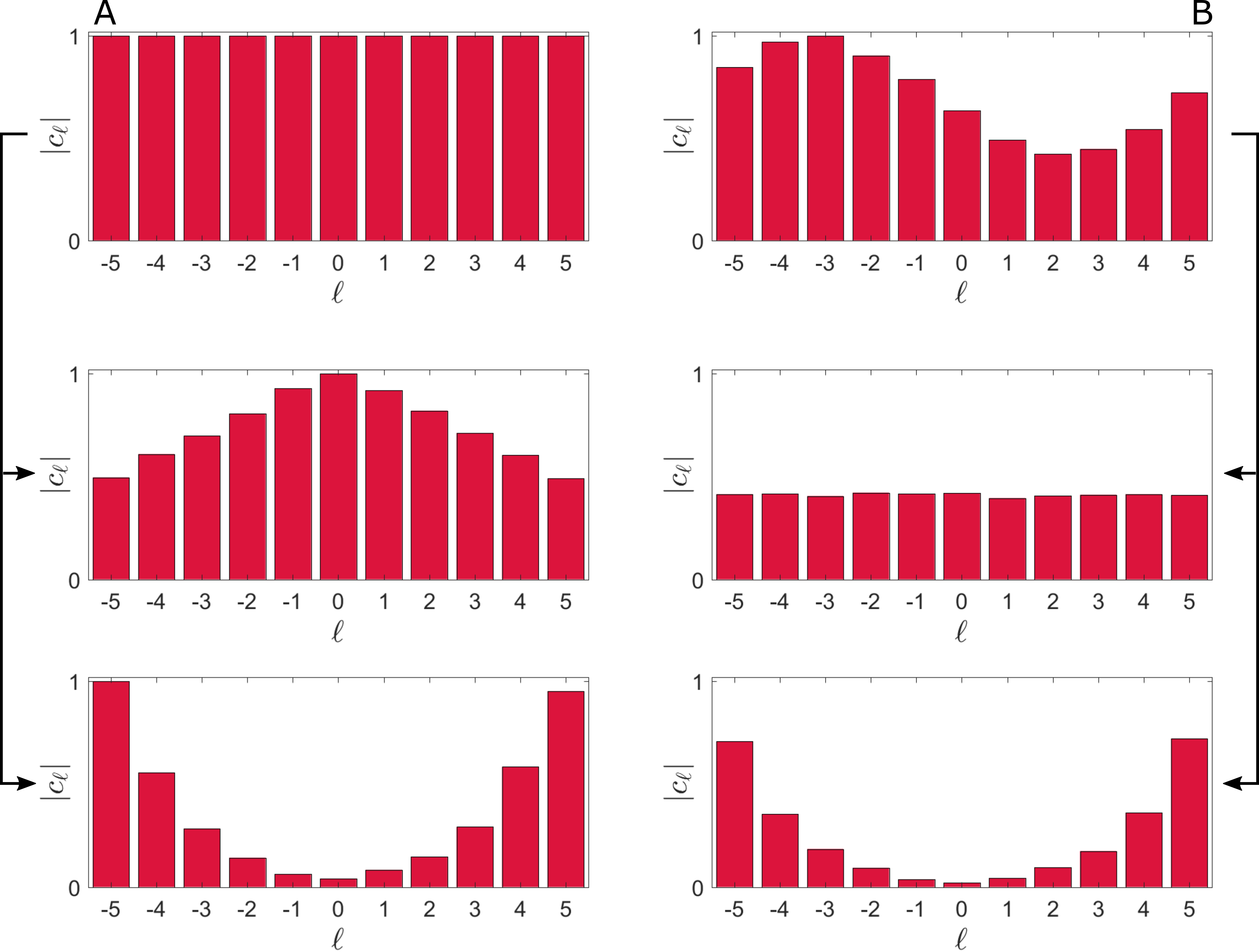}
\caption{Experimental results showing examples of the shaping of a flat (A) and non-flat (B) OAM spectrum. All spectra have been corrected in accordance with the procedure in Fig.~\ref{fig:corrections} and all $c_\ell$ values are normalised to the maximum camera gray level.}
\label{fig:shaping}
\end{figure*}

\begin{figure*}[ht]
\centering
\includegraphics[width = \textwidth]{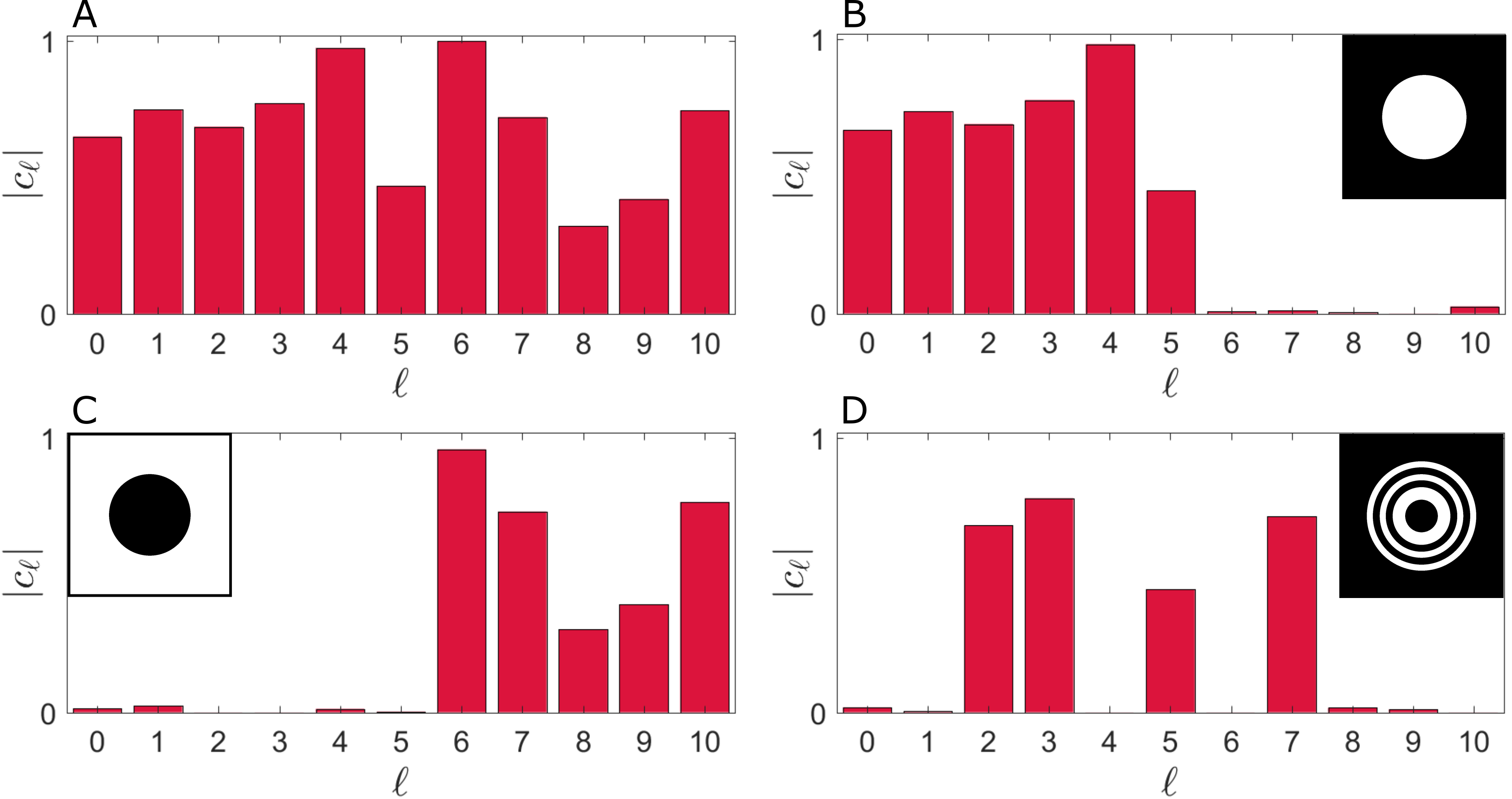}
\caption{Experimental OAM filtering: A is the initial OAM spectrum and B, C and D are the spectra after low-pass, high pass and prime-pass filtering, respectively. Insets show the corresponding OAM filter holograms.}
\label{fig:filter}
\end{figure*}

Once the setup is calibrated and the correction factors found, more exotic OAM shaping can be pursued. We show four such examples in Fig.~\ref{fig:shaping}: where an initial uniform spectrum (A) is transformed to a triangular-shaped spectrum and a U-shaped spectrum and where an initial sinusoidal spectrum (B) is transformed to a flat spectrum and a U-shaped spectrum, all in a single step. Here, the initial spectrum is known and $|t_\ell|$ was determined by dividing the desired coefficients by the initial coefficients. Figure \ref{fig:filter}, on the other hand, shows three examples of shaping where the initial spectrum is not known. Here, a random OAM spectrum was generated and three different OAM filters were applied: a low-pass, high-pass and prime-pass filer.

In principle, there are no limitations to how the shaping can be done, one need only adjust the transmission function of the ring apertures accordingly, remembering that the amplitude of spectrum coefficients can only be made smaller in our implementation.

\subsection*{Inter-Modal Phase Shaping}
Although we do not explicitly show inter-modal phase shaping in our previous examples, the procedure is very similar to how the amplitude of $c_\ell$ is manipulated and is straightforward to implement if utilising SLMs. The difference is that $t_\ell$ becomes complex and so one also encodes a constant phase value $\arg\,(t_\ell)$ within the ring aperture which is defined relative to an arbitrary reference mode in the spectrum.

\begin{figure*}[h!] 
	\centering
	\includegraphics[width=\textwidth]{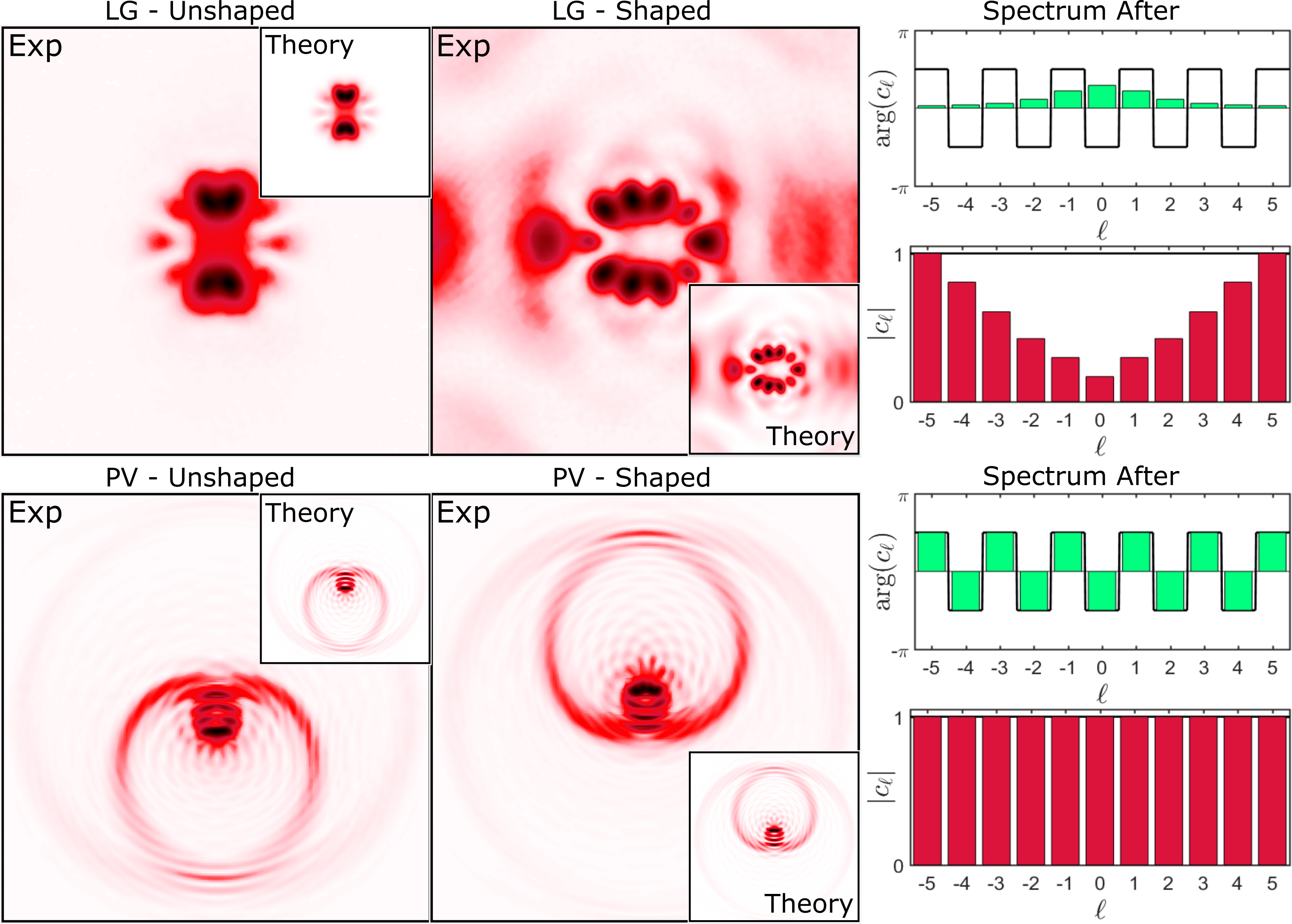}
	\caption{Demonstration of the shaping of inter-modal phases for a uniform OAM superposition in $\ell \in [-5,5]$. The camera images are taken at the Fourier plane where the phase manipulations are clearly visible. Simulated OAM spectra are shown alongside where the dark lines correspond to the encoded ring aperture transmission functions. The fact that the experimental field intensities are in such good agreement with simulation gives qualitative evidence that the experimental OAM spectrum has been shaped in accordance with the simulated spectrum.}
	\label{fig:Phases}
\end{figure*}

To showcase this, we present qualitative evidence (instead of doing the full modal decomposition) that our scheme has shaped the inter-modal phases. In this regard, we experimentally generated the uniform LG and PV superpositions as in Fig.~\ref{fig:modeOverlap} (having no inter-modal phases) and encoded $\arg(t_\ell) = (-1)^{|\ell|-1} \pi/2$ and $|t_\ell| = 1$ within each ring aperture. Since the amplitude of the field is unaffected at the modification plane, we can observe the phase changes at the Fourier plane. The results are shown in Fig.~\ref{fig:Phases}. The insets correspond to simulated intensities at the Fourier plane having the OAM spectrum as shown alongside the figure. We argue that the high fidelity between the theoretical and experimental images is qualitative evidence for the shaping of inter-modal phases.

Similar to the shaping of $|c_\ell|$ for LG modes, the spatial overlap between LG-OAM modes inhibits the ability to modify the inter-modal phases independently. Thus, it is difficult to determine the exact relationship between $\arg(t_\ell)$ (what we encode) and $\arg(f(\ell))$ (the transformation we want). In addition, shaping the inter-modal phases also has an effect on the amplitude of $c_\ell$ for the case of LGs; phase discontinuities at the ring aperture boundaries invariably affect the amplitude of the beam. For PVs, on the other hand, the engineered spatial independence of each OAM mode facilitates $\arg(f(\ell)) = \arg(t_\ell)$ and the amplitude of the OAM spectrum is unchanged.

\section{Perspectives}
Overall, our OAM shaping protocol from start to finish involves 3 steps: mapping OAM states to specific spatial positions, constructing binary ring apertures for those regions and then modifying the amount and phase of light within those regions. Step 1 is specific to the particular vortex modes chosen and step 3 depends on how one wishes to shape the OAM spectrum; these two steps are not subject to much change. Of the three, step 2 is the most malleable. If the OAM modes are sufficiently spatially separated (as is the case with the engineered PV superposition) then choosing ring apertures in such a way as to enclose the entire mode is the most fruitful approach: there is a direct equivalence between the desired shaping (defined by $f(\ell)$) and what is encoded on the ring aperture (defined by $t_\ell$). With regards to ``regular" vortex modes, perhaps future work could explore different ways of constructing the ring apertures than the approach we considered here. It may be that there is a more optimal choice which will enable the shaping of these vortex modes to be more predictable.

Since it is significantly more difficult to add/inject light into specific annular regions, this meant that we only considered the case where the amplitude of the OAM coefficients is lessened. One way to increase the relative contribution of a particular OAM mode is to decrease the relative contribution of all other modes. This is equivalent but has the potential disadvantage of significantly reducing the overall power in the beam. A solution to this would be to utilise a gain medium in conjunction with a SLM. In this way, individual modes in the OAM spectrum could be both raised and lowered in amplitude.


\section{Conclusion}
To summarise, we proposed and demonstrated a scheme for single-step modulation of a light field's OAM spectrum which utilises a quasi-mapping between the position and OAM spaces. This grants one the ability to shape the OAM spectrum using conventional devices (such as SLMs). Using our approach, we have shown that complete control of the amplitude and phase of the OAM spectrum is achievable in a single step. While most vortex mode sets have an inevitable spatial overlap, PV modes can be engineered with minimal spatial overlap, thus facilitating greater control over the OAM spectrum. Since OAM is being utilised more and more for applications in classical and quantum information theory, we conjecture that the ability to arbitrarily modify the OAM spectrum could drive new techniques that would enhance some of these technologies. 




\bibliography{mypaperdatabase}


\end{document}